# The electromagnetic fields under, on and over Earth surface as "when, where and how" earthquake precursor

Strachimir Chterev Mavrodiev, mavrodi@inrne.bas.bg
INRNE, BAS, Sofia, Bulgaria

**Abstract**

It is given an attempt for statistical estimation of "when" earthquake prediction for Balkan- Black Sea region using the geomagnetic field signal.

The preliminary test of the approach for England (Hartland), Turkey (Kandilli) and India (Alibag) regions is presented.

The step by step research program for creating **"when, where and how" earthquake prediction region Network system** on the basis of the experimental data for geomagnetic field, electro-potential distribution in the Earth crust and atmosphere, temperature Earth crust distribution, gravitational anomaly map, season and day independent temperature depth distribution, water sources parameters (debit, temperature, chemical composition, radioactivity), gas emissions, ionosphere condition parameters, Earth radiation belt, Sun wind, crust parameters (strain, deformation, displacement) and biological precursors is proposed.

The achievements of tidal potential modeling of Earth surface with included ocean and atmosphere tidal influences, many component correlation analysis and the nonlinear inverse problem methods in fluids dynamics and Maxwell equations are crucial.

The today almost real time technologies GIS for archiving, analysis, visualization and interpretation of the data and non- linear inverse problem methods for building theoretical models for the parameter behaviors, correlations and dynamics have to be used.

**Introduction**

The "when, where and how" earthquake prediction is not a solved problem [1].

The more precision space and time set for Earth's crust condition parameters and the including in the monitoring the electromagnetic fields measurements under, on and over Earth surface, the temperature distribution and other possible precursors can be useful for research of the "When, where and how" earthquake's prediction [2]. The progress in this direction is summarized in [3].

In Part 1 are given the 2002 statistics estimations for the reliability of the time window earthquake prediction on the basis of geomagnetic field measurements and Earth tidal behavior [3] for Balkan, Black Sea region.

In Part 2 the preliminary analysis of the approach is applied for England, Turkey an India region.

In Part 3 the list of monitoring parameters, which can be useful, the theoretical apparatus for analysis and technology for data acquisition are described shortly

**Part 1. The geomagnetic field likes a time window precursor for Balkan, Black Sea region-2002**

The one geomagnetic vector projection GMF is measured with accuracy less then or equal to1 nT (detector: know-how of JINR, Dubna, Boris Vasiliev) with 2.4 samples per second, Nm is equal to 144 samples per minute.

The minute averaged value $GMF_m$ and its error $\Delta GMF_m$ are given by

$$GMF_m = \sum_{i=1,Nm} GMF_i / Nm \text{ and } \Delta GMF_m = \sum_{i=1,Nm} \Delta GMF_i / Nm.$$

The standard deviation $\sigma GMF_m$ and its error $\sigma \Delta GMF_m$ are

$$\sigma GMF_m = (sqrt (\Sigma_{i=1, Nm} (GMF_i / GMFm-1)^2)/ Nm ,$$

$$\sigma \Delta GMF_m = (sqrt(\Sigma_{i=1,Nm} (\Delta GMF_i / \Delta GMF_m -1)^2)/ Nm.$$

After some time (from 1999 to 2001) of looking for correlations between the behavior of the geomagnetic field, Earth tidal gravitational potential and the occurred earthquakes one turn out that the daily averaged value of $\sigma GMF_m$ ($\sigma \Delta GMF_m$), which we denote by Sig ($\Delta$Sig), is playing the role of earthquake precursor.

The next Figure 1 illustrates the behavior of geomagnetic field and its variation for a day without signal for near future "big and near enough" earthquake in the region.

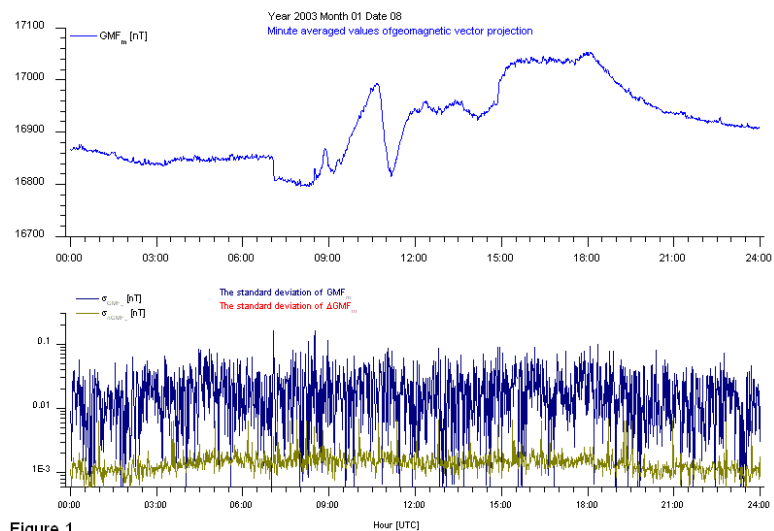

Figure 1
The behavior of geomagnetic field for a normal day- without a signal for a near future earthquake

The Figure 2 illustrates the behavior of geomagnetic field and its variation for a day with a signal for near future earthquake. One has to be sure there are not a cosmos reasons for the geomagnetic quake. So, for example, see the sites [8].

The time window of the incoming events is defined by the next date of the Earth tidal potential extremum with tolerance +/-1 day in the case of minimum and +/-2 days in the case of maximum. For the example in Figure 2 the predicted time was at June 4 +/- 1 day.

The answer of the question for distinguishing the predicted event (or group of events – aftershocks) from the events which can occur in the region in the same time window is given by the parameter SChtM:

$$SChtM = Mag / ((R_{EQ} + Distance) / 1000)^2,$$

where [Distance] = km, $R_{EQ}$ [km] is the radios of earthquake volume. Roughly, at this stage of our knowledge, we accept $R_{EQ}$ = 30 km.

The parameter SChtM is a measure of the energy influence of the earthquakes at the device point.

For the concrete example in Figure 2 the predicted earthquake had occur with parameters: Time 6/4/2002 21:10, Lat.: 40.45, Long.: 21.75, Dep.: 27, Mag.: 3.1, Dist.: 271 and SChtM: 42.

At this stage of research all earthquakes have the same SChtM parameter for different definitions of the Magnitude. After developing on the basis of inverse nonlinear problem the empirical and theoretical dependences between incoming earthquake processes, magnetic quake and parameters of earthquake we will arrive to a set of SChtM parameter in correspondence with the different definition of Magnitude. The volume, its depth, chemical and geological structures have to be included in the dependences.

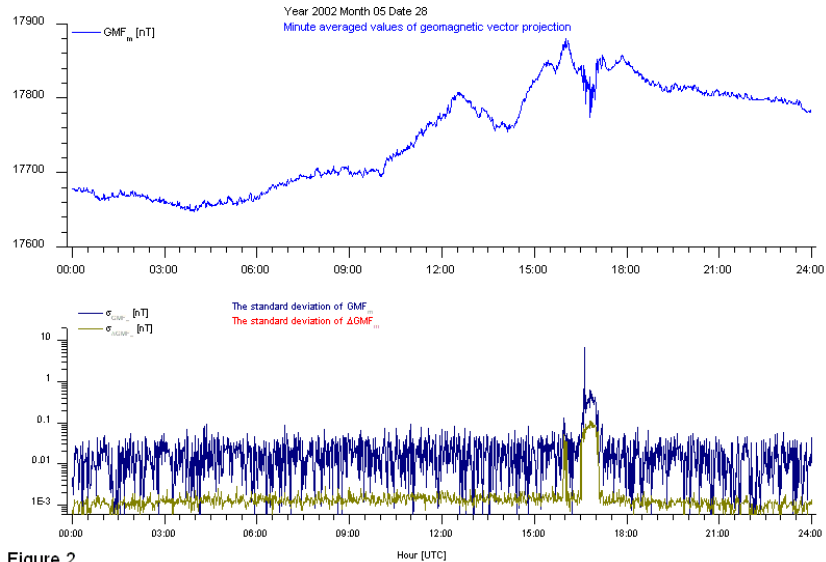

Figure 2
The behavior of geomagnetic field for a day with a signal for a near future earthquake

It is interesting to stress that in the case of big earthquake with Mag > 6, in more than 60 % of the cases, after the earthquake there are very little variations of the magnetic field with different time duration: from 10 minute to some hours. Those influences do not depend on the distances, but one can see the differences, which depend on the zone of earthquake: convergence or divergence one. See for example next Figure 3.

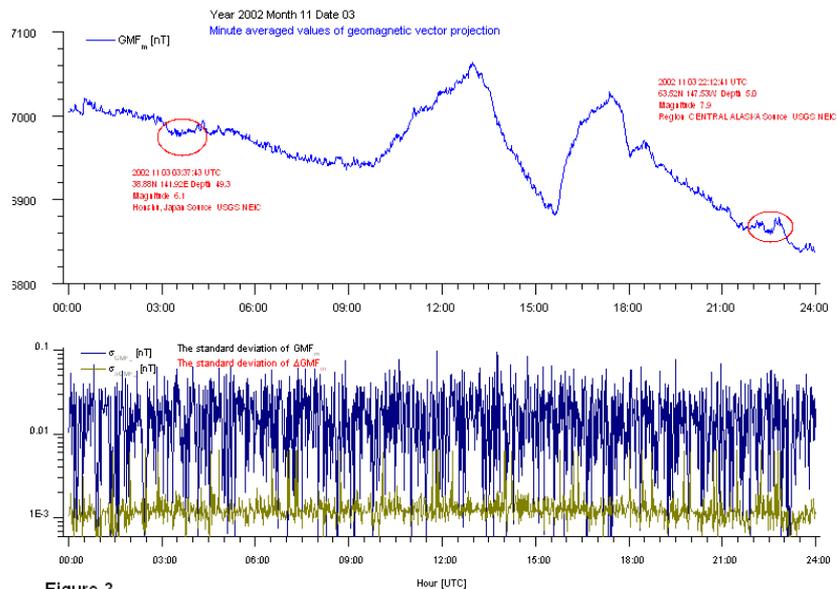

Figure 3
A registration of big world earthquake, Mag>6

In the next Table 1 are represented 2002 data for dates of magnetic quake and of predicted time window for 37 near future earthquakes.

|  | Table 1 | EQ Predicted | time |  |  |  |  |
|---|---|---|---|---|---|---|---|
| Number | Signal Time DDMMYY | Predicted Time DDMMYY | Time Window [Day] | Number | Signal Time DDMMYY | Predicted Time DDMMYY | Time Window [Day] |
| 1 | 27/12/2001 | 06/01/2002 | 1 | 22 | 30/06/2002 | 02/07/2002 | 2 |
| 2 | 17/01/2002 | 21/01/2002 | 1 | 23 | 04/07/2002 | 09/07/2002 | 1 |
| 3 | 27/01/2002 | 02/02/2002 | 1 | 24 | 11/07/2002 | 16/07/2002 | 2 |
| 4 | 31/01/2002 | 09/02/2002 | 2 | 25 | 15/07/2002 | 22/07/2002 | 1 |
| 5 | 08/02/2002 | 17/02/2002 | 1 | 26 | 21/07/2002 | 31/07/2002 | 2 |
| 6 | 18/02/2002 | 25/02/2002 | 2 | 27 | 03/08/2002 | 03/08/2002 | 1 |
| 7 | 28/02/2002 | 03/03/2002 | 1 |  |  | 13/08/2002 | 1 |
| 8 | 08/03/2002 | 17/03/2002 | 1 |  |  | 20/08/2002 | 2 |
| 9 | 18/03/2002 | 27/03/2002 | 4 | 28 | 26/08/2002 | 28/08/2002 | 2 |
| 10 | 03/04/2002 | 08/04/2002 | 1 | 29 | 07/09/2002 | 10/09/2002 | 2 |
| 11 | 08/04/2002 | 16/04/2002 | 2 | 30 | 10/09/2002 | 16/09/2002 | 1 |
| 12 | 15/04/2002 | 23/04/2002 | 1 | 31 | 20/09/2002 | 25/09/2002 | 1 |
| 13 | 23/04/2002 | 28/04/2002 | 2 | 32 | 25/09/2002 | 08/10/2002 | 1 |
| 14 | 01/05/2002 | 06/05/2002 | 1 |  |  | 16/10/2002 | 1 |
| 15 | 09/05/2002 | 14/05/2002 | 2 | 33 | 12/10/2002 | 25/10/2002 | 2 |
| 16 | 18/05/2002 | 20/05/2002 | 1 |  |  | 06/11/2002 | 2 |
| 17 | 23/05/2002 | 26/05/2002 | 2 | 34 | 07/11/2002 | 13/11/2002 | 2 |
| 18 | 28/05/2002 | 04/06/2002 | 1 | 35 | 12/11/2002 | 21/11/2002 | 1 |
| 19 | 08/06/2002 | 11/06/2002 | 2 | 36 | 22/11/2002 | 28/11/2002 | 2 |
| 20 | 15/06/2002 | 18/06/2002 | 2 | 37 | 06/12/2002 | 12/12/2002 | 1 |
| 21 | 21/06/2002 | 24/06/2002 | 1 |  |  |  |  |

In the time period because of hardware problems a signals for the geomagnetic quake was not published four times.

The independent control (In the framework of Strasbourg recommendations about earthquake prediction of the European Union for ethical and public security reasons) of the prediction reliability was organized in the framework of the Bulgarian Academy of Sciences, its Geophysical Institute and a set of colleagues, which are interested in the topic or research.

So, the time of 37 events was predicted successfully, four was let pass, because of hardware problems. The notion "event" means one or more earthquakes with the same epicenter and in some times 2 or 3 earthquakes, which occurred in the same time window, but in different epicenters. The main reasons for this ambiguity are:
1. The measuring of only one (instead of geomagnetic vector) geomagnetic projection in one point (at last 2, better 3);
2. The absence of data under analysis for electro-potential distribution in Crust.

In the next Figure 4 are presented the distribution of the time deviation (the difference between predicted time and the time of occurred earthquake) for all predicted earthquakes and its Gauss fit.

**The good Gauss fit of the distribution, the growth of the Gauss fit amplitude and constant behavior of Gauss fit parameter w have to be considered like prove of reliability of the time window prediction basis on the data for geomagnetic quake and Earth tidal behavior.** The error with one day will be improved after including in the analyzed data the hourly tidal behavior.

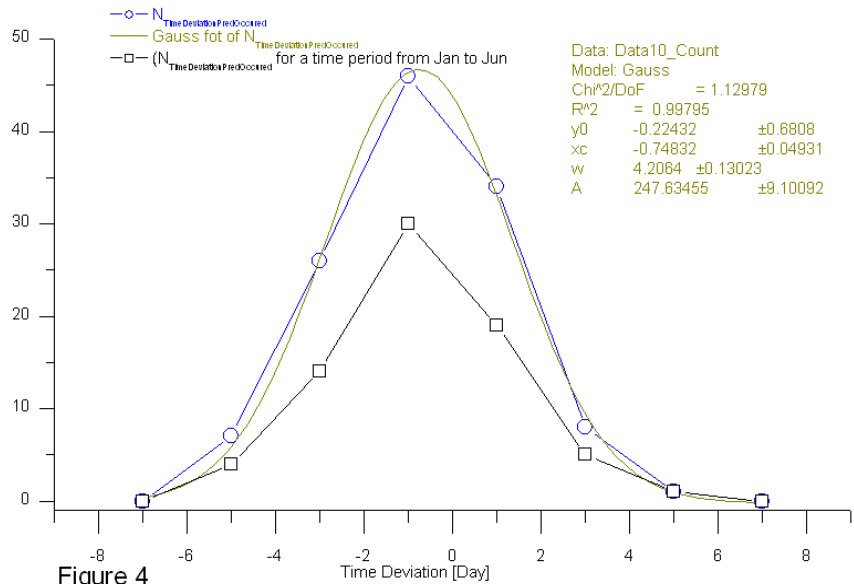

Figure 4
The time deviation distribution (predicted- occured)

In the Figure 5 are presented for a comparison the distributions of the magnitude for all earthquakes in the region and for predicted one. It is seen the character of distributions is the same, which is implementer argument that the approach is based on the real physical dependences.

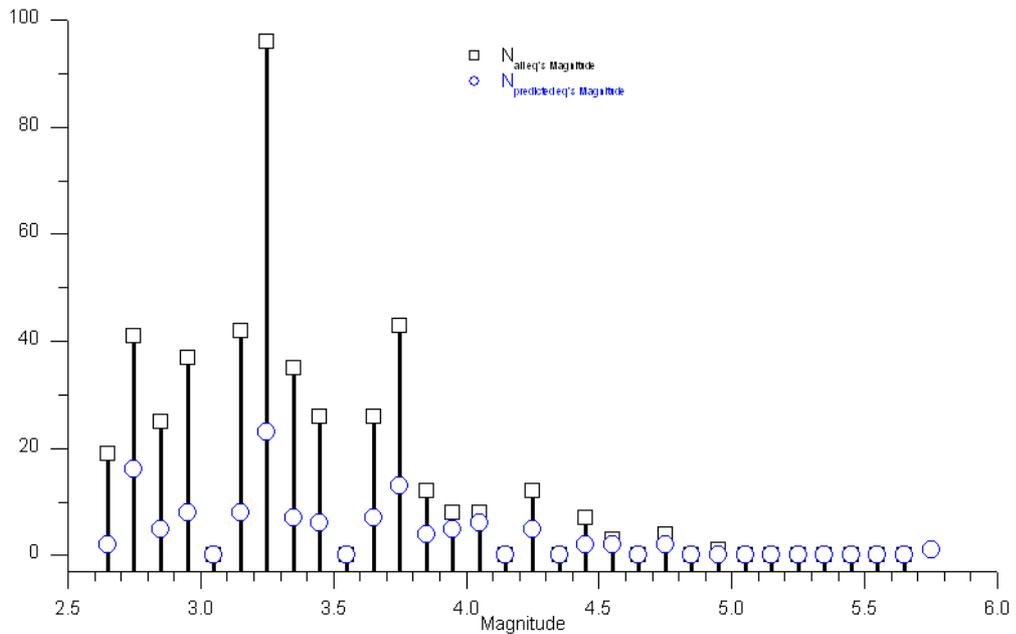

Figure 5
The magnitude distribution for all and predicted earthquakes

The next Figure 6 is the 2002 NEIC earthquakes map for the region.

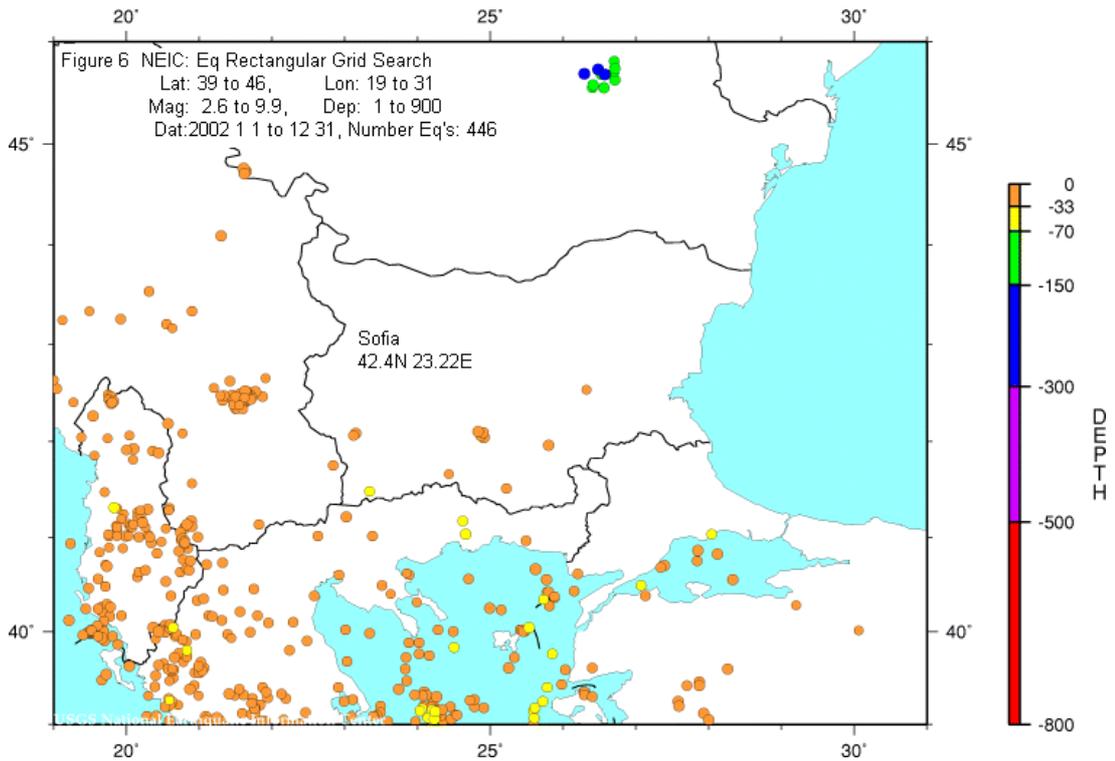

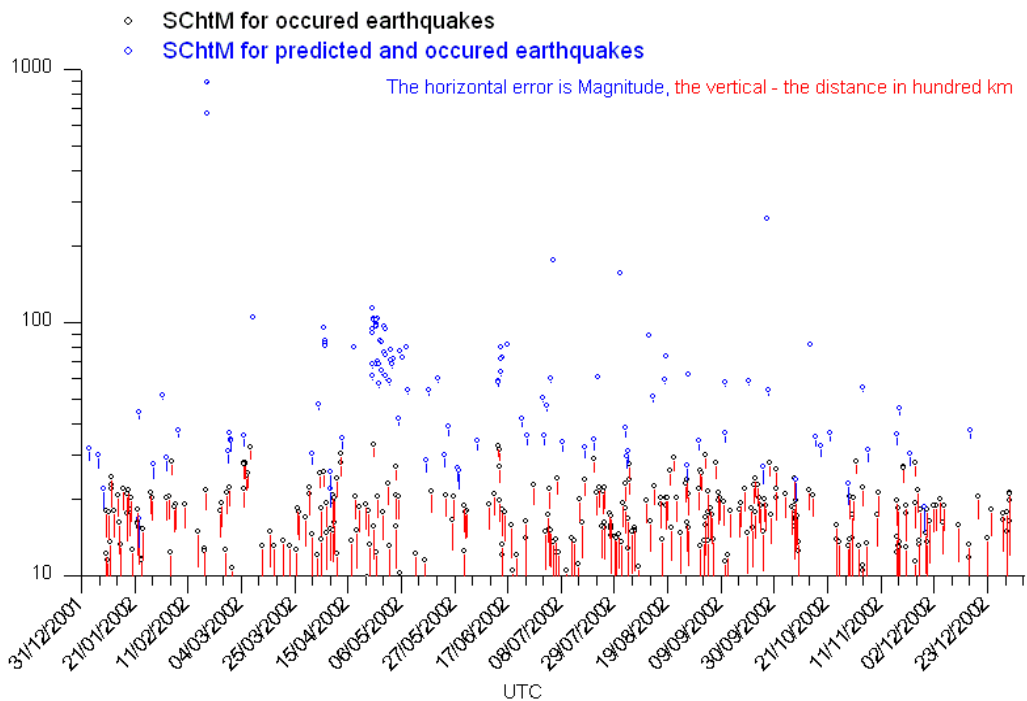

Figure 7
The SChtM parameter for all occured earthquakes

The next Figure 8 illustrates the distances to which the time window for future event can be predicted in the framework of the approach. One must stress that the aria is not symmetric one, which is connected with the geological history of the region.

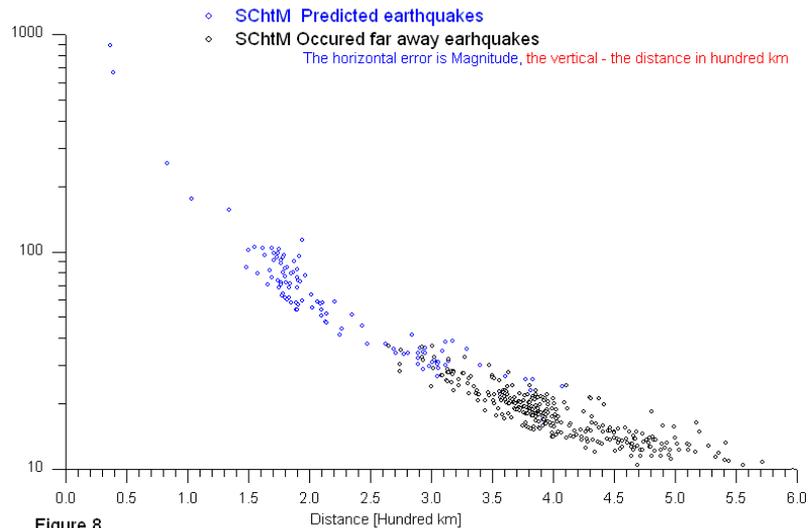

Figure 8
The SchtM parameter for all 2002 earthquakes as function of distance

In conclusion one can say that the earthquake time prediction approach on the basis of accurate measuring of the one geomagnetic vector projection can be accepted to work well at distance to 250 km, not uniquely to 450- 500 km, for SChtM > 20 and magnitude interval 2.6<Mag<5.3. From occurred 41 events, 37 were predicted and 4 was let pas. From 446 occurred earthquakes, the epicenters of 122 earthquakes were in the predictable region.

**Part 2. The application of the approach for England, Turkey an India**

**On the 23 September 2002 England 4.8 Earthquake**
The next Figure 9 represents the EQ,' map for Hartland Geomagnetic station region for the period January to September, 2002.

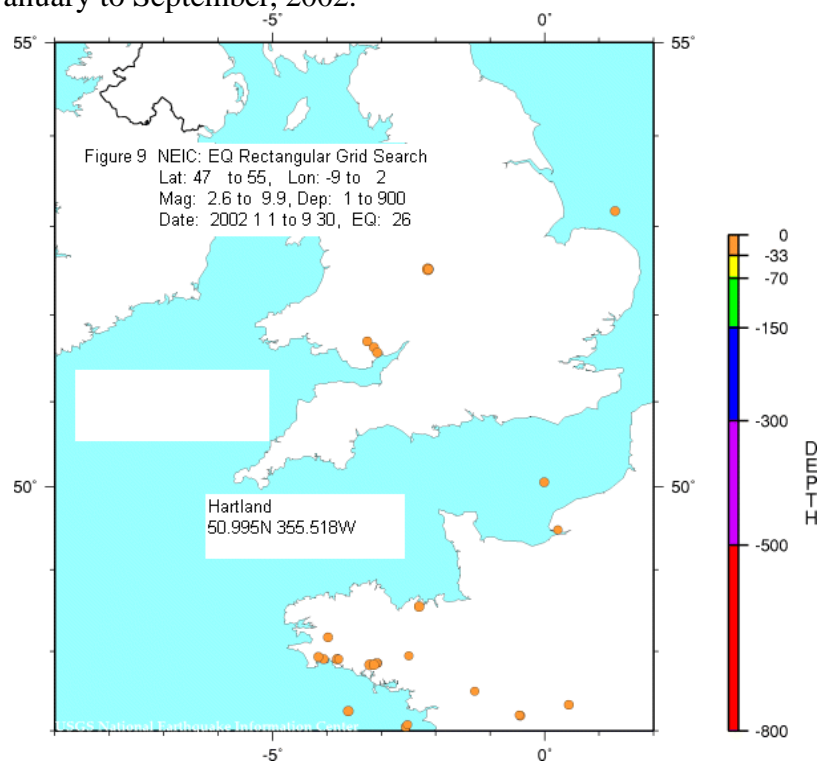

The Figure 10 represents the correspondence between the SChtM earthquake parameter and preceding maximum of the geomagnetic signal

$$Sig = sqrt\,(\sigma_H^2 + \sigma_D^2 + \sigma_Z^2\,),$$

where $\sigma_H$, $\sigma_D$ and $\sigma_Z$ are the standard daily deviation of hourly data for H, D and Z.

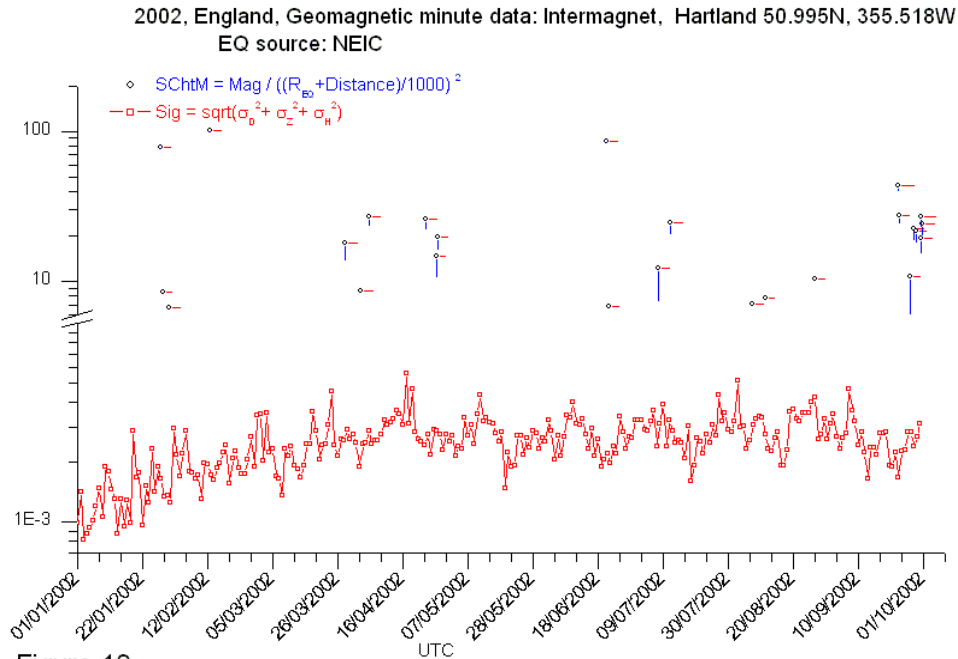

Figure 10
The geomagnetic quake as earthquake precursor

The correspondence is not unique for all earthquakes. There are some peaks of the geomagnetic signal without earthquakes in the next 14 day period. But for earthquakes with SChtM >20 the correspondence is like for the Balkan, Black Sea region.

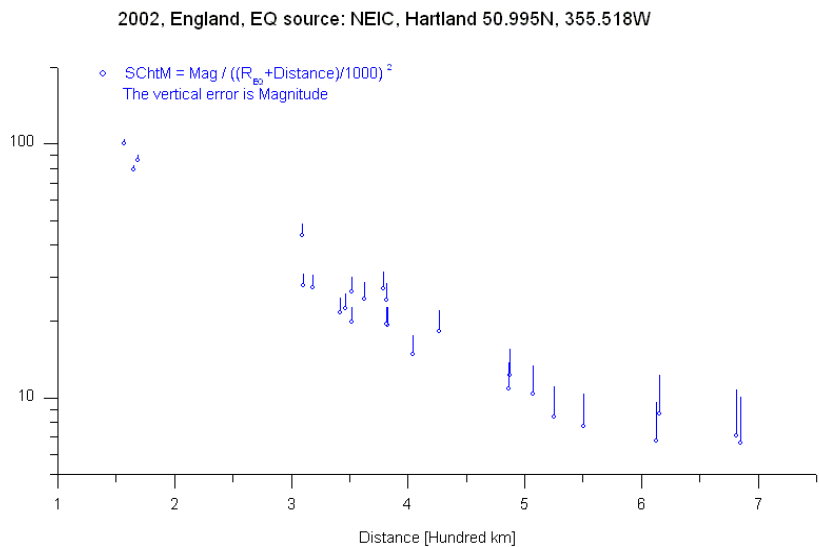

Figure 11
The SChtM eq parameter as function of distance

From Figure 11 one can see the SChtM dependence from the distance for occurred in 2002 earthquakes in the Hartland region.

**The geomagnetic quake before earthquake with parameters 22/07/2002 05:45:03.02, Lat.:50.89, Lon.6.10, Dep.:17, ML5.20, Station GRF appeared at 07/09/2002 HDZ Hartland measurements**.

The samples of the measurement, the record time, the different dependences of the signal from standard deviations of the geomagnetic vector components and influence of ocean tidal influence has to be carefully analyzed.

**Turkey 2002**

The Figure 12 represents the eq's NEIC map for Istanbul region- Kandilli geomagnetic Observatory with 41.07N, 29.06E coordinates.

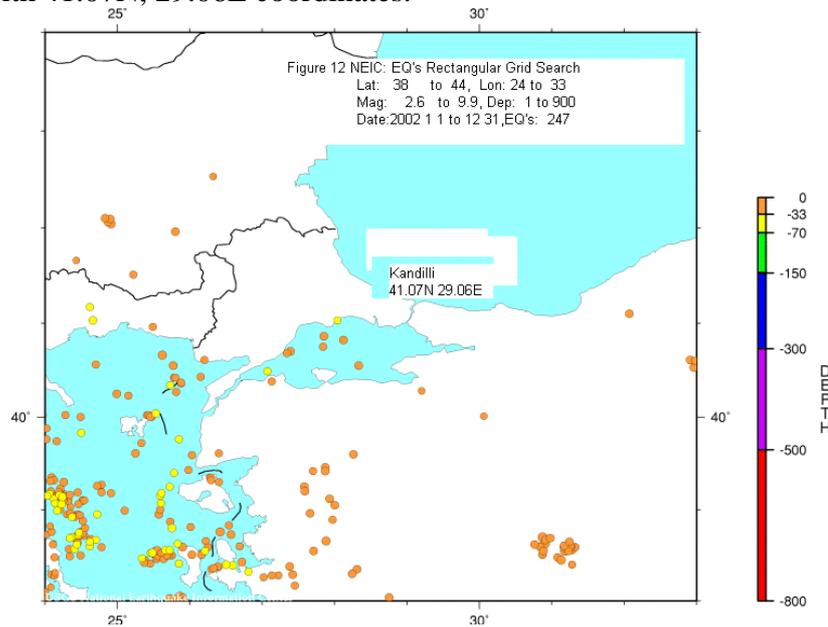

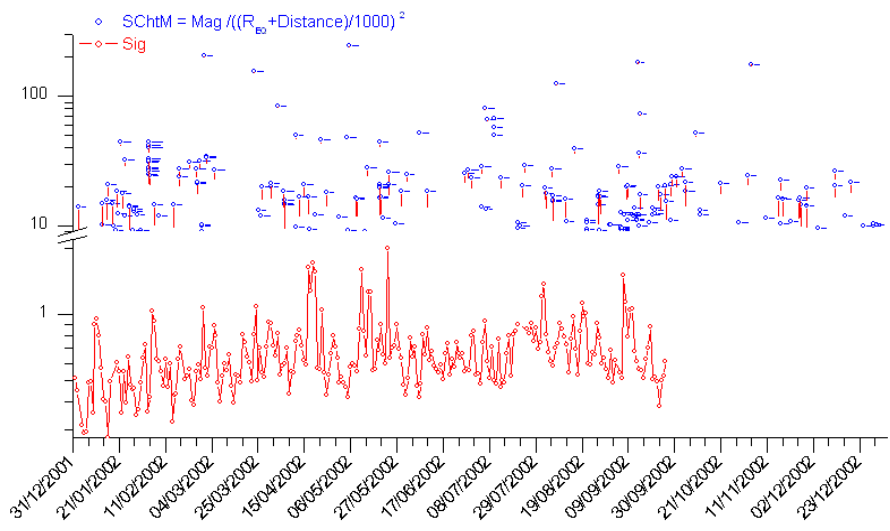

Figure 13
The geomagnetic quake as earthquake precursor

From the next Figure 13 is seen the correspondence between the time of the earthquakes with big SChtM parameter and the foregoing geomagnetic quake in 14 days time window.

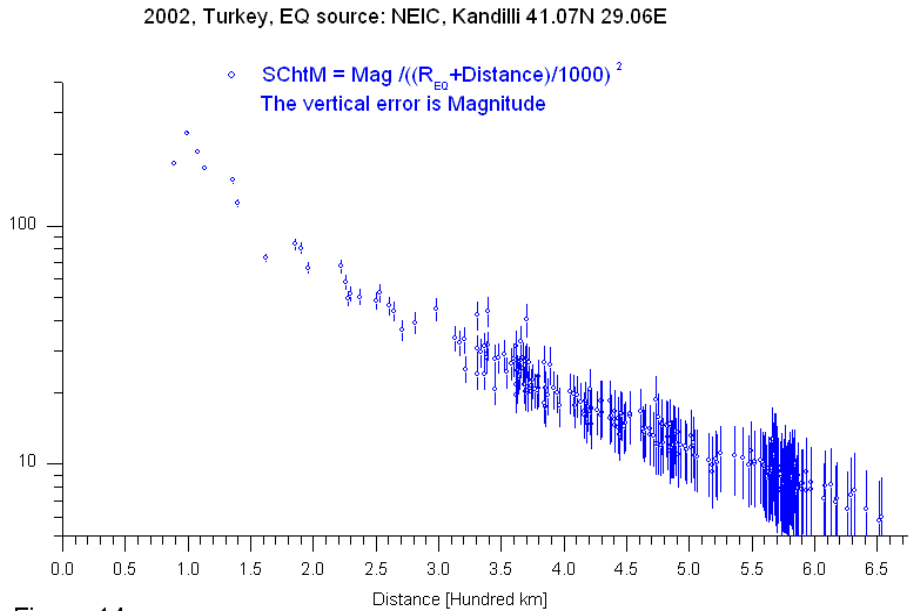

Figure 14
The SChtM eq parameter as function of distance

As for Sofia data the SChtM parameter has the same behavior with distance.

### India 2001

In the next Figure 15 are represented the earthquake in India region (Alibag Geomagnetic Observatory, 18,63N, 72,76E). It is seen the Gujarat events from 26 January, 2001

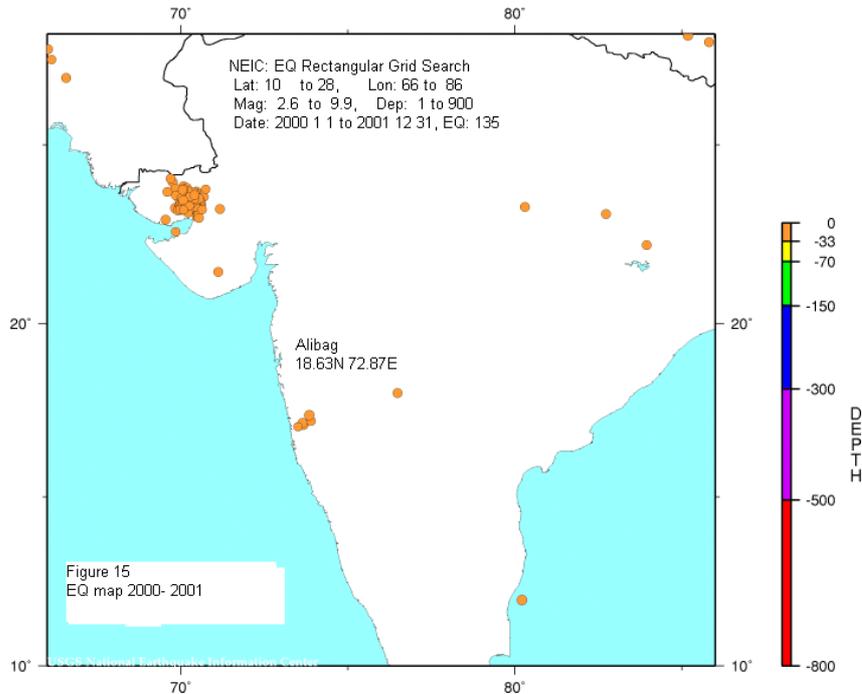

From Figure 16 is seen the distribution of the earthquakes magnitude for the analyzed period.

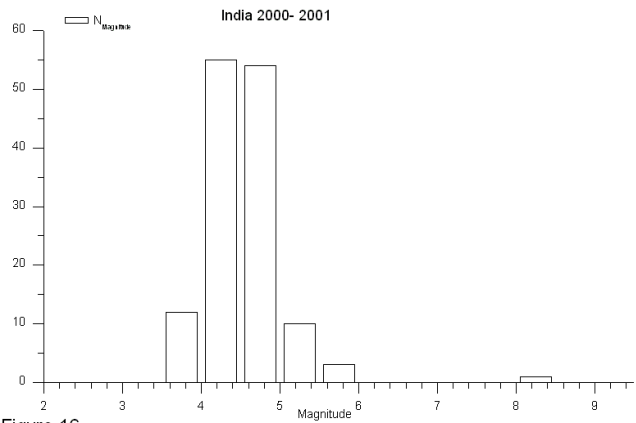

Figure 16
Distribution of the earthquakes magnitude

From Figure 17 is seen how faraway are the earthquakes risk zone from the Alibag Observatory: 200 km, 600 km, 1200 km and 1800 km.

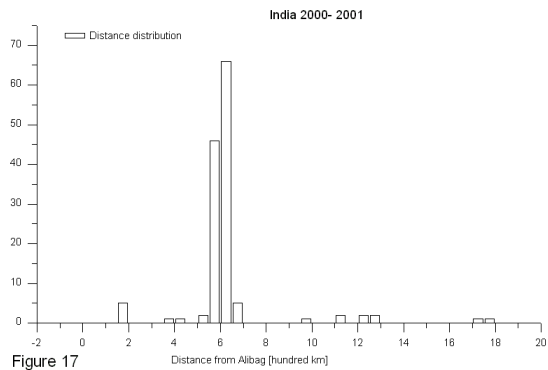

Figure 17
EQ's Distance from Alibag Distribution

In Figure 18 is represented the correspondence between the earthquakes and geomagnetic quakes for 2000.

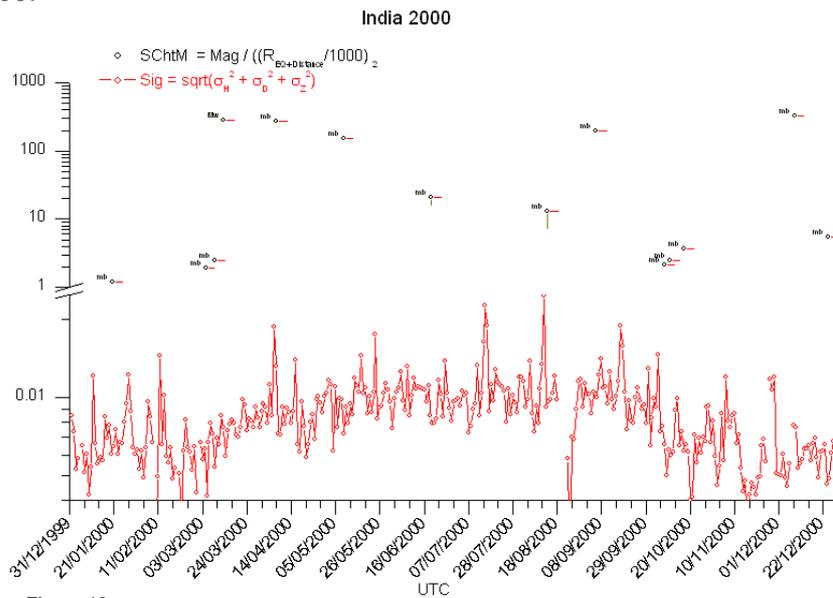

Figure 18
SChtM paramater and geomagnitic signal as earthquake precursor

The next Figure 18 illustrates the reliability of the approach for predicting the Gujarat big 8.2 earthquake on the basis of geomagnetic signal estimated from minute data for F.

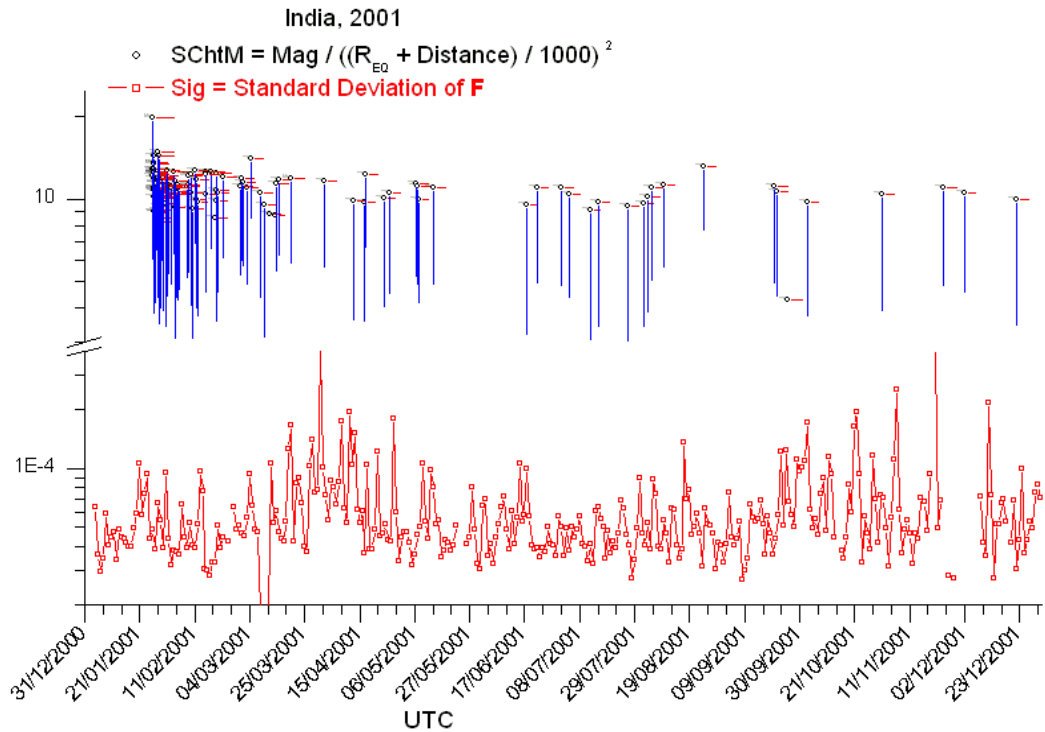

Figure 18 SChtM eq's parameter and geomagnetic signal

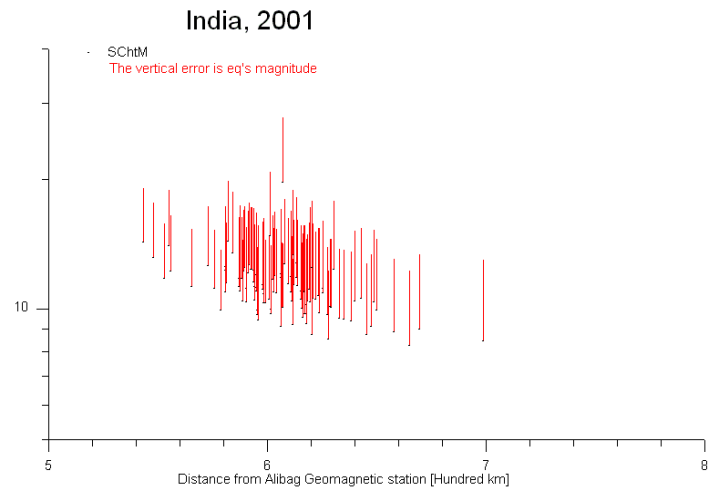

Figure 19 SChtM eq's parameter as finction of distance

In Figure 19 the SChtM dependence is represented. It is seen well the main event and the next aftershocks.

The samples of the geomagnetic measurements, the record time, the different dependences of the signal from standard deviations of the geomagnetic vector components and the set of Geomagnetic Observatories have to be carefully analyzed for real application of the proposed approach fro the region.

**Part 3. Proposal for creating of Short Time Earthquake prediction local NETWORK**

Hear we will not discuss the long time prognostic system for estimation of earthquake risk. They are well known [10].

The aim of this Proposal is to create a system for research the reliability of the local forecast system for earthquakes in the interval Mag > 2.6- 3 and radius till 600 km. The system is complex and the attended practical result will be an adequate physical theoretical model for the Earth magnetism. The system includes experimental, theoretical and technological parts:

*Experimental data*
1. *Geomagnetic field,*
2. *Electro-potential distribution in the Earth crust and atmosphere,*
3. *Temperature Earth crust distribution,*
4. *Crust parameters (strain, deformation, displacement)*
5. *Gravitational anomaly map,*
6. *Season and day independent depth temperature distribution,*
7. *Water sources parameters (debit, temperature, chemical composition, radioactivity),*
8. *Gas emissions,*
9. *Ionosphere condition parameters,*
10. *Infrared radiation of Earth surface, earthquake clouds*
11. *Earth radiation belt,*
12. *Sun wind,*
13. *Biological precursors.*

*Theory*
1. *The achievements of tidal potential modeling of Earth surface with included ocean and atmosphere tidal influences,*
2. *Many component correlation analysis*
3. *Nonlinear inverse problem methods in fluids dynamics and Maxwell equations are crucial.*

*Technologies*

*GIS for archiving, analysis, visualization and interpretation of the data and non-linear inverse problem methods for building theoretical models for the parameter behaviors, correlations and dynamics.*

The set of the devices has to be in correspondence with known data for earthquakes risk zone (gravitational anomalyties and Crust parameters monitoring (strain, deformation, displacement). The geomagnetic device set distance has to be in order of 300 – 600 km, the electro- potential 100- 300 km in dependence of geological today situation and its history. The set for monitoring of the daily and season crust temperatures has to be in order 300 km. The correlations with Sun wind influence have to be in real time.

The system has to be created step by step. The condition for next step pass has to be the building of physical clear new theoretical correlations or dynamical models and, of course, the successful: "when", "when, where" or "when, where and how" earthquake prediction.

**14 days window "when" prediction**

For short time "when" prediction the set of geomagnetic monitoring with the accuracy, accepted from Intermagnet [9], but with 50 samples per second for H, D, Z or X, Y, Z component measurements and 1 sample per second for **F.** The minute averaged standard deviation has to be calculated in real time. The electro- potential distribution dynamics under, on and over Earth surface will correlates with geomagnetic data. The estimation of the epicenter coordinates from 3 points geomagnetic devices on the basis of inverse problem is possible. The time is fixed from the next minimum (+/-1 day time tolerance) or maximum (+/-2 days) of daily averaged tidal potential. The influence of the ocean and atmosphere tides has to be under consideration.

### 14 days window "when, where" prediction

The epicenter coordinates of near future event has to be estimated by the analysis of geomagnetic end electro-potential data. The information for the daily and season independent temperature distribution, infrared activity, ionosphere parameters, gravitational anomaly set, the crust parameters dynamics and the hazard risk estimation for the region, has to improve the coordinate's estimation. In the case of big events the appearing of "earthquakes clouds" (Zhonghao Shou) and quasi magnetic pole in the region, which can be fixed from the radiation belts changes, will be important.

### 14 days window "when, where, how" prediction

The gradient distributions of the above mentioned parameters, the burned new, more adequate physical models for parameter correlations and dynamics will give a data for step by step statistical estimation of the reliability of prediction.

## Conclusion

The solution of **"when, where, how" earthquake prediction problem** obviously needs the creation of very big science group. The presence in the group of the "skeptics" is very important.

## Acknowledgements

The author is very thankful to Prof. Ramesh P. Singh for invitation to present a paper to this Workshop.

The citation list is not full as it had to be and the author is waiting the contacts with the colleagues for apologizing and further collaboration.

## References:

1. **Skeptics**:
   http://earthquake.usgs.gov/hazards/prediction.html
   Predicting earthquakes, Louis Pakiser and Kaye M. Shedlock, USGS;
   Is the reliable prediction of individual earthquakes a realistic scientific goal?, Debate in *NATURE, 1999,;*
   Earthquake prediction information, R. Ludwin, University of Washington,
   Research activities at Parkfield, California,
   Assessment of schemes for earthquake, Royal Astronomical Society Meeting Abstracts,
   Earthquake prediction, societal implications, K. Aki, Univ. Southern California, From *Reviews of Geophysics,*
   Earthquakes cannot be predicted, Robert J. Geller, D. D. Jackson, Y. Y. Kagan, F. Mulargia, From *SCIENCE;*
   Papers from an NASA Colloquium "On Earthquake prediction: The Scientific Challenge"
2. **Optimists**:
   http://www.cosis.net/members/frame.php?url=www.copernicus.org/EGS/EGS.html, 27th General Assembly, Acropolis, Nice, France, Natural Hazards, NH10, Seismic hazard evaluation, precursory phenomena and reliability of prediction, Contadakis M., Biagi P., Zschau J., April 2002,

    Zhonghao Shou, http://quake.exit.com/,

    R. Dean, www.earthquakeforecast.org,

    C.Thanassoulas, http://www.earthquakeprediction.gr/,

    J. Tsatsaragos, http://users.otenet.gr/~bm-ohexwb/alert2.htm,

    B. Ustundag, http://www.deprem.cs.itu.edu.tr/

3. S.Cht.Mavrodiev, C.Thanassoulas, "Possible correlation between electromagnetic earth fields and future earthquakes", INRNE-BAS, Seminar proceedings, 23- 27 July, 2001, Sofia, Bulgaria, ISBN 954-9820-05-X, 2001, http://arXiv.org/abs/physics/0110012,

    S.Cht.Mavrodiev, The electromagnetic fields under, on and up Earth surface as earthquakes precursor in the Balkans and Black Sea regions, www.arXiv.org, physics, Subj-class: Geophysics; Atmospheric and Oceanic Physics, http://arXiv.org/abs/0202031, February, 2002:

4. S.Cht.Mavrodiev, On the short time prediction of earthquakes in Balkan- Black Sea region based on geomagnetic field measurements and tide gravitational potential behavior, http://arXiv.org/abs/physics/0210080, Oct, 2002;

5. http://wwwneic.cr.usgs.gov;

6. http://www.geomag.bgs.ac.uk/gifs/on_line_gifs.html;

7. Venedikov A.P., Arnoso R., Vieira R., A Program for tidal data processing, Computer&Geoscience, in press, 2002,

    Venedikov A., Arnoso R., "Program VAV/2000 for Tidal Analysis of Unevenly Spaced Data with Irregular Drift and Colored Noise", J. Geodetic Society of Japan, vol.47, 1, 281- 286; 2001.

8. http://www.sec.noaa.gov/SWN/

   http://www.sec.noaa.gov/rt_plots/satenv.html

   http://www.sec.noaa.gov/rt_plots/xray_5m.html

9. Intermagnet, http://obsmag.ipgp.jussieu.fr/cgi-bin/form,

    http://swdcwww.kugi.kyoto-u.ac.jp/wdc/Sec3.html

10. V. I. Keilis-Borok, http://www.mitp.ru